\documentclass[aps,twocolumn,prc,floatfix,showpacs,superscriptaddress]{revtex4-1}
\usepackage{graphicx}
\usepackage{subfigure}
\usepackage{color,colordvi}
\usepackage{dcolumn}
\usepackage{amsmath}
\usepackage{amssymb}
\usepackage{float}
\usepackage{epstopdf}
\usepackage{placeins}
\usepackage{ulem}  
\newcommand{\ts}{\textstyle }
\renewcommand{\ts}{\textstyle }
\newcommand{\etal}{\textit{et al.}}
\begin{document}

\author{Wytse van Dijk}
\affiliation{Department of Physics, Redeemer
University College, Ancaster, ON, Canada L9K 1J4} \affiliation{Department of Physics and Astronomy, McMaster University,   Hamilton, ON, Canada L8S 4M1}
\email{vandijk@physics.mcmaster.ca}
\author{Calvin Lobo}
\affiliation{Department of Physics and Astronomy, McMaster University,   Hamilton, ON, Canada L8S 4M1}
\author{Alison MacDonald}
\affiliation{Department of Physics, University of Alberta, Edmonton, AB, Canada T6G 2E9} 
\author{Rajat K Bhaduri}
\affiliation{Department of Physics and Astronomy, McMaster University,   Hamilton, ON, Canada L8S 4M1}

\title{Fisher zeros of a unitary Bose gas}

\begin{abstract}

For real inverse temperature $\beta$, the canonical partition
function is always positive, being a sum of positive terms. There are
zeros, however, on the complex $\beta$ plane that are called Fisher zeros. 
In the thermodynamic limit, the Fisher zeros coalesce into
continuous curves. In case there is a phase transition, the zeros
tend to pinch the real-$\beta$ axis. For an ideal trapped Bose gas in an isotropic three-dimensional harmonic oscillator, this tendency is clearly seen, signalling Bose-Einstein condensation (BEC).  The calculation can be formulated exactly in terms of the virial expansion with temperature-dependent virial coefficients. When the second   virial coefficient of a strongly interacting attractive unitary gas
  is included in the calculation, BEC seems to survive, with 
the condensation temperature shifted to a lower value for the unitary gas. This shift is consistent with a direct calculation of the heat capacity from the canonical partition function of the ideal and the unitary gas.

\end{abstract}

\maketitle

\section{Introduction}
The canonical partition function $Z_N(\beta)$ of a $N$-particle system is expressed as
a function of $\beta$, the inverse of the temperature $T$ in units of $k_B$, the Boltzmann constant. For real $\beta$, it
is always positive, being a sum positive terms. For complex $\beta$,
however, the zeros come in complex conjugate pairs on the
complex-$\beta$ plane. In the thermodynamic limit, these zeros coalesce into
continuous curves. Fisher~\cite{fisher65} pointed out that in the event of a phase
transition, the zeros pinch the positive real axis  at a value of $\beta$  
 that corresponds to
the transition temperature. In this paper, we first consider ideal
bosons trapped in an isotropic harmonic oscillator. 
This ideal case was first studied by M\"ulken \etal~\cite{mulken01},
who were interested in the order of the phase transition.  
Our focus in this paper is on the changes in this pattern of Fisher zeros when a strong attractive interaction between the atoms is introduced. In particular, does BEC survive in this situation? 

While the pattern of the Fisher zeros of a unitary bose gas has not been studied earlier, a large number of papers
have studied the stability of trapped bosons with attractive
interaction~\cite{{dalfovo99}, {cornish00},{rem13},
{fletcher13},{li12},{makotyn14}}.
For a bose gas, there is loss of atoms through three-body
recombination: $A+A+A\rightarrow A_2+A$ ~\cite{esry99,esry06}. 
The number of three-body
recombinations per unit volume per unit time is proportional to $C(a)
n^3 a^4$, where $a$ is the scattering length, $n$ the number density
of atoms, and $C(a)$ is a dimensionless $a$-dependent constant.  Near unitary strength, Efimov  trimers are formed, causing atomic
losses that increase rapidly with increasing $a$. The equilibration
rate, however, scales as $n a^2 v$, where $v$ is the average
speed. Thus it would appear that the system would lose the atoms
before equilibration. Makotyn \etal~\cite{makotyn14}
realized, however, that for a gas at unitarity, the
scattering length drops out, and the only length scale is the average 
interparticle spacing, that goes like  $n^{-1/3}$. Invoking
universality, the energy varies as $n^{2/3}$, and the time scales like $n^{-2/3}$.  
The  equilibration {\it rate}, as well as the loss {\it rate} of a unitary Bose gas
should both go like $n^{2/3} $. Which of these will dominate has to be
determined experimentally.
One is ignoring here the inhomogeniety of the gas due to initial trapping,
and assuming the range of the force to be too small to be relevant. 
Makotyn \etal~\cite{makotyn14} start  with a BEC condensate of 
$^{85}$Rb atoms far from Feshbach resonance, 
and then adjust  the magnetic field to rapidly  attain the unitary
limit. Subsequently the gas is moved away from resonance, and
allowed to evolve ballistically. They find
that the gas equilibrates rapidly compared to the loss time. While
they cannot tell whether the BEC survives in the unitary gas, they
find  the momentum distribution of the atoms to be remarkably similar to  that of the ideal degenerate Bose gas.     
Their work demonstrates that the unitary Bose gas can be prepared and dynamically investigated.  

Recently Piatecki and Krauth~\cite{piatecki14} have performed a many-body
 Monte-Carlo calculation to describe the Efimov-driven phase
 transition of a unitary bose gas. Their Hamiltonian not only consists
 of a two-body attractive zero-range potential, but also a three-body 
hard core cut-off in the hyper radial coordinate. This prevents a
Thomas collapse of the trimers. The bosons are
trapped in an isotropic harmonic oscillator potential. 
The full equation of state consists of three phases: the unitary  bose
gas in the normal phase, a superfluid bose condensate, and a
superfluid of Efimov trimers.    Piatecki and Krauth  compute the
equation of state in the gas phase  numerically and in their
  Fig.~2 compare it with the theoretical result including the second
and third virial coefficients. The agreement is good up to
  slightly above the BEC transition temperature.
The bose condensate at unitarity that they find is also robust, with the condensate temperature somewhat lower than the ideal case. We shall discuss this later after presenting our results.  

In this paper, we consider the Fisher zeros of both the ideal and the
unitary bose gas. This requires the analytical form of the $N$-body 
canonical  partition function $Z_N(\beta)$. This is known
exactly for the ideal bose gas through poperly defined temperature
dependent statistical virial coefficients (Sect 3.1).  
For the ideal gas, taking into  account a large number of statistical
virial coefficients, both sides  of the peak of the heat capacity at $T_c$ are clearly
obtained. This suggests that the  virial expansion, with proper
temperature dependence retained, is  applicable deep in the degenerate
regime.  Unfortunately we cannot follow the same procedure in detail for the unitary gas since only the second virial coefficient is known analytically.   
The third virial coefficient for a homogeneous gas has been calculated in \cite{castin13_fr} by Castin and Werner.  However, it involves three-body Efimovian parameters in an essential way, and is given analytically only at the low- and high-temperature ends. 
Excluding the third and higher virial coefficients is a limitation of our approach, 
since we do not expect the condensate temperature to be accurately  
described then. Nevertheless, 
 we shall see how the pattern of the Fisher zeros changes due to the
 interaction at the two-body cluster level with a corresponding
 shift in the condensation temperature. This description allows us to 
see if BEC survives the strong attractive interaction at unitarity. 
Our calculation of the pattern of Fisher zeros qualitatively suggests that BEC survives at unitarity, and the pattern is  somewhat similar to  that of the Fisher zeros of an ideal Bose gas.  Since we do not include the third virial coefficient, we have no knowledge of the Efimov driven transition. 

It should be noted that in a discussion of a unitary fermion
gas~\cite{ku12,bhaduri12,bhaduri13} it is shown that a
high-temperature virial expansion can match the equation of state over
a large range of fugacity, but not
all the way to the superfluid phase.  In this paper we explore similar
behaviour for BEC by studying the  Fisher zeros.  

The notion of unitarity in lower spatial dimensions have been examined Castin and Werner~\cite{castin12}, and by How and LeClair~\cite{how10}.  For one-dimensional bosons, modelling the interaction by a delta function interaction $g\delta(x)$, unitarity may be defined by the $S$-matrix being equal to $-1$, and is obtained by $g\rightarrow \infty$.  This impenetrable bose gas  is equivalent to a spinless free fermi gas~\cite{girardeau60}.  In two dimensions, unitarity may be defined by the scattering length $a_{2D}\rightarrow \infty$, but this maps onto the free space problem.

The two-dimensional Bose gas trapped in an harmonic oscillator and interacting with a weakly repulsive force was examined semiclassically in~\cite{bhaduri00,bhaduri02a} and more rigorously by Gies \etal~\cite{gies04}.  In~\cite{gies04} it was shown that long wavelength fluctuations are suppressed by the HO trap, and BEC is well defined for a zero-range repulsive interaction.

\section{Partition Function}
The canonical partition function (for fixed $N$) is defined  as 
\begin{equation}
Z_N(\beta)=\sum_{E_i^{(N)}} \exp\left(-\beta E_i^{(N)}\right)~,
\label{eq:01}
\end{equation}
where  $E_i^{(N)}$ are the complete set of
eigenenergies of the $N$-body system including states in the continuum, if any. The sum is taken over all states
${i}$ including the degenerate ones. Since the energies $E_i^{(N)}$
depend on the volume of the system, there is a  volume 
dependence in $Z_N(\beta)$.  
To appreciate the connection between complex zeros near the real axis
to critical fluctuations at phase transition, let us consider
the canonical partition function $Z_N(\beta)$ on the complex-$\beta$ plane. This may be written as a product of $\widetilde{Z}(\beta)$ with no complex zeros and factors explicitly showing the zeros, 
\begin{equation}
Z_N(\beta)=\widetilde{Z}_N(\beta)\prod_r
\left(1-\frac{\beta}{\beta_r}\right)\left(1-\frac{\beta}
{\beta_r^{*}}\right)~.
\label{fish} 
\end{equation}
For finite boson number $N$, the index $r$ runs over a finite number 
${\cal N}(N)$, where ${\cal N}$ is much larger than $N$. 
The pairs of complex conjugate zeros ensures that the partition
function remains real on the real-$\beta$ axis.
Knowing the distribution of zeros, one may, in principle, calculate the thermodynamic properties of the system.

The grand canonical partition function, on the other hand, allows for 
particle exchange (in addition to energy) via the reservoir, and is defined as 
\begin{equation}
\begin{split}
{\cal Z}(\beta, z) & =\sum_{N=0}^{\infty}\sum_{E_i^{(N)}}
 \exp\left(-\beta E_i^{(N)}+\beta \mu N\right) \\ 
 & =\sum_{N=0}^{\infty} Z_N(\beta) z^N~,
\label{eq:02}
\end{split}
\end{equation}
where the fugacity $z=\exp (\beta \mu)$ and $Z_0(\beta)$ is defined to
be one. We shall use this form presently to include virial
coefficients in the recursion relation to calculate the $N$-boson canonical partition functions.  

\FloatBarrier 

\section{ Trapped Bosons and BEC}
\subsection{Calculation of $ Z_N(\beta)$}
\label{next}

\begin{figure}[t]
\centering                                                                                      
\caption{The Fisher zeros for a systems of 50 or 100 ideal trapped bosons taking into account the exact discrete energy spectrum.
}
\label{fig:01}$\begin{array}{c@{\hspace{0.0in}}c@{\hspace{0.0in}}}                                                                                                             
\resizebox{3.5in}{!}{\includegraphics[angle=-90]{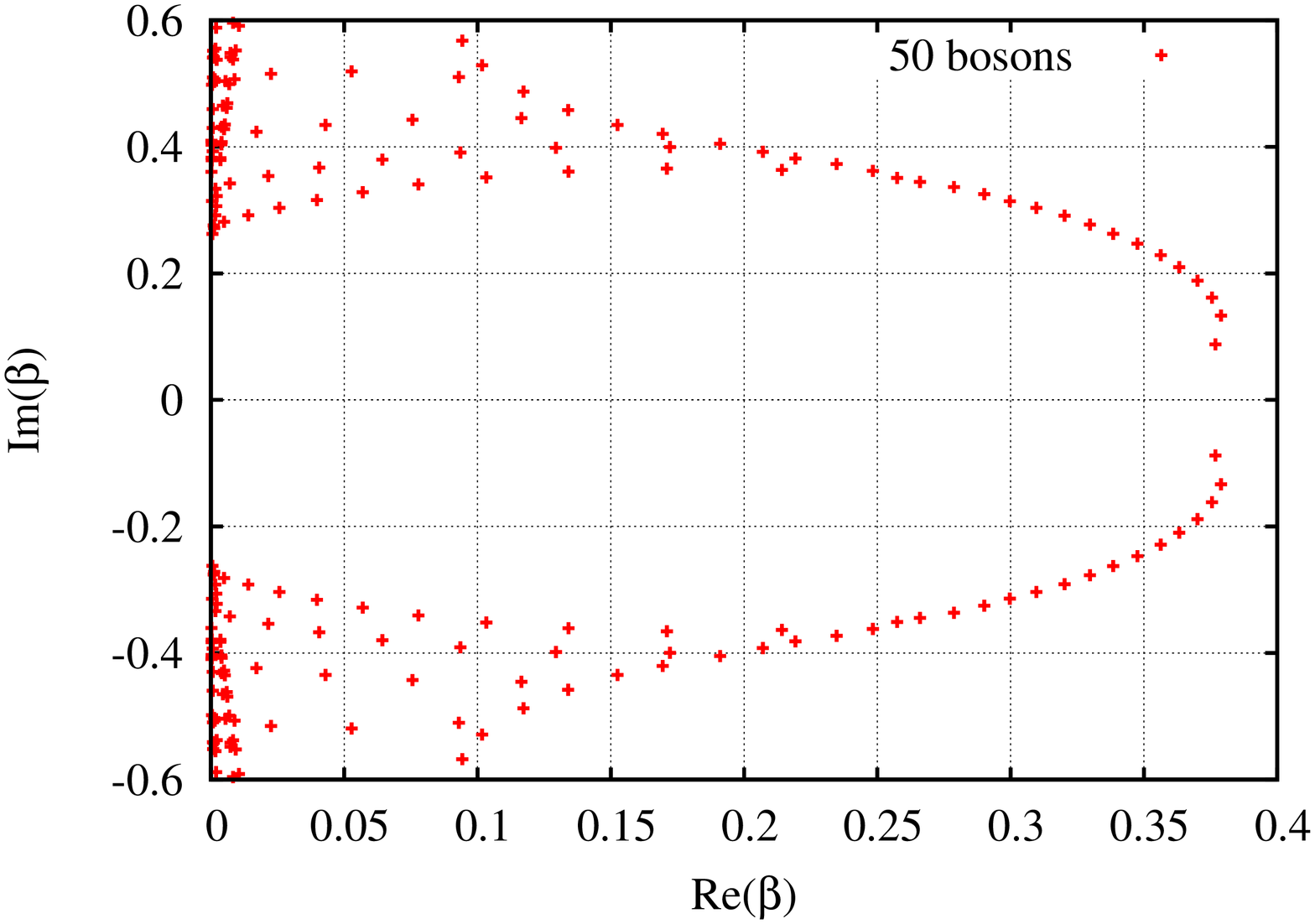}}  \\ 
\resizebox{3.5in}{!}{\includegraphics[angle=-90]{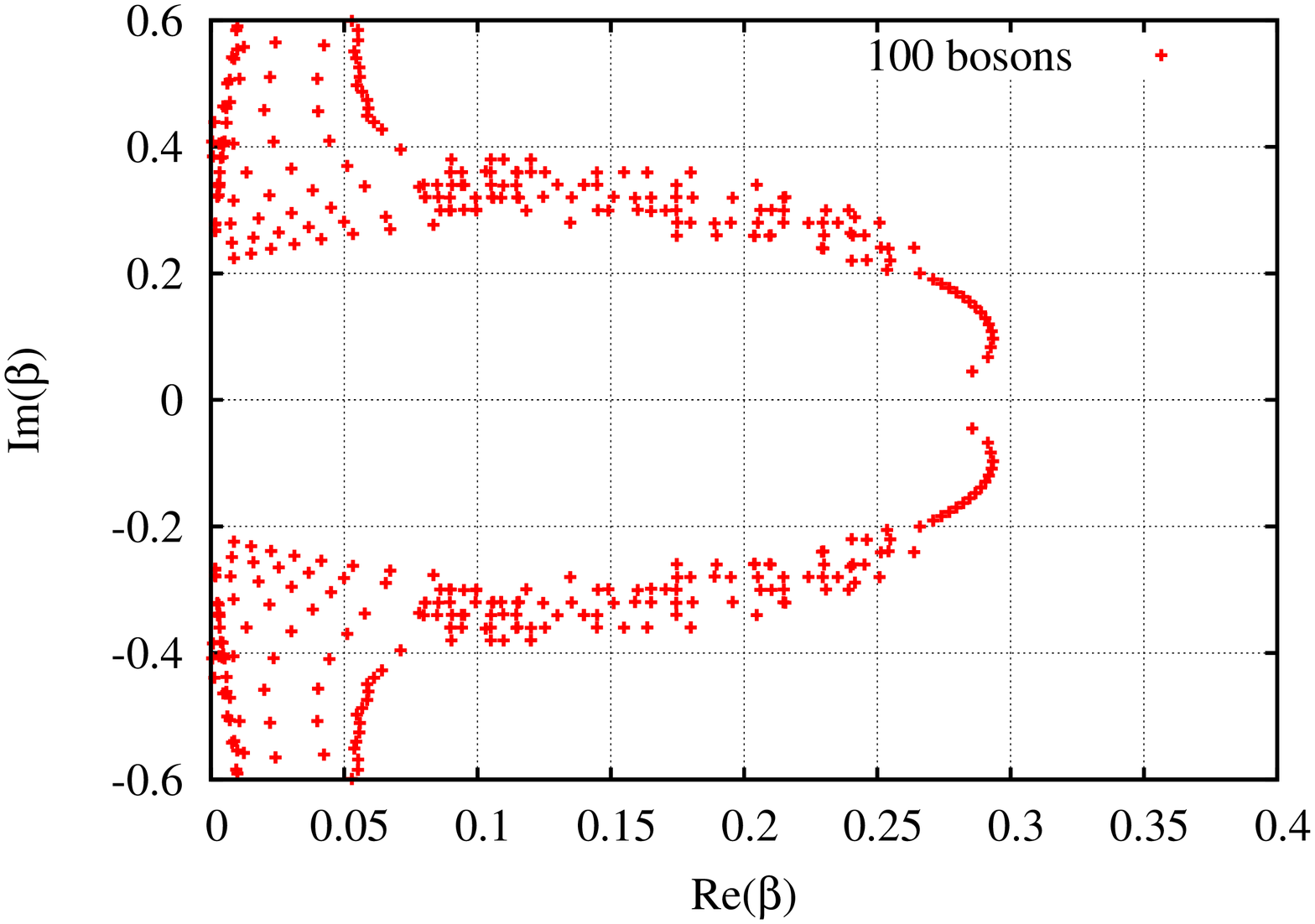}}
\end{array}$
\end{figure}

In order to obtain the Fisher zeros~\cite{fisher65}, 
we need to calculate the canonical partition function $Z_N(\beta)$. 
This in turn requires an analytical form for the one-body canonical partition
function $Z_1(\beta)$.
For a single particle
 in a three-dimensional harmonic oscillator, this is given by     
\begin{equation}\label{kish}
 Z_1(\beta)=\left(\sum_{n=0}^\infty e^{\displaystyle
     -\beta\varepsilon_n}\right)^3 =\left(\dfrac{e^{\ts
       -\beta\hbar\omega/2}}{1-e^{\ts -\beta\hbar\omega}}\right)^3, ~
\end{equation} 
where $\varepsilon_n=(n + 1/2)\hbar\omega$. This is the exact quantum 
$Z_1(\beta)$. The often-used continuous version of $Z_1(\beta)$ is the 
high-temperature limit of (\ref{kish}) :  
\begin{equation}
Z_1(\beta)=\dfrac{1}{(\hbar\omega\beta)^3}~.
\label{conti}
\end{equation}
It  gives a continuous density of single-particle states that grows as $\epsilon^2$.  In subsequent calculations we put $\hbar\omega=1$.

For an ideal boson gas, $Z_N(\beta)$ may be obtained from $Z_1(\beta)$ using a
recursion relation~\cite{borrmann93, mullin03}. For bosons,  
\begin{equation}
Z_N(\beta)=\frac{1}{N}\sum_{k=1}^{N}Z_1(k\beta) Z_{N-k}(\beta)~,
\label{class}
\end{equation}
where $Z_0(\beta)=1$. Using the recursion relation above,
$Z_2(\beta)=(Z_1^2(\beta)+
Z_1(2\beta))/2$. Similarly, $Z_3(\beta)$ is expressed in terms of
$Z_1(\beta) $ and $Z_2(\beta)$, and so on.
Typically, in a magnetic trap, $N$ is very large, of the order of $10^6/$cc or larger for an oscillator with $\hbar\omega \simeq 1$ nK. For our calculation, it is
impractical to take such large $N$ in the recursion relation.  
Theoretically, for harmonically trapped bosons, BEC is manifest even for a small number of
bosons~\cite{mulken01,ketterle96}, though the transition temperature
is not so sharp. 
 So far as Fisher zeros are concerned, they still
approach the real-$\beta$ axis, but stop short of it (see Fig.~\ref{fig:01}).

\begin{figure}[ht]
\centering  
\caption{The virial coefficients of the ideal bose gas as  functions of $\beta$.} 
\label{fig:02}
\resizebox{3.5in}{!}{\includegraphics[angle=-90]{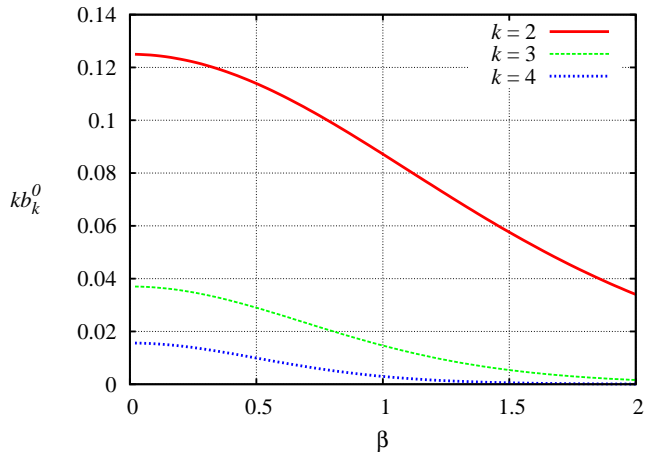}}                                                                                    
\end{figure}

It is also straightforward to include the effect of interaction in the
 unitary gas at temperatures spanning the degenerate regime, as
was done for the fermionic gas in \cite{ku12,bhaduri12, bhaduri13}.
This is done by 
using virial cluster formalism in the grand canonical partition
function. Note that 
\begin{equation}
\ln {\cal Z}(\beta,z)=Z_1(\beta) \sum_{l=1}^{\infty} b_lz^l~,
\label{eq:8}
\end{equation} 
where $b_l$ is the $l^{th}$ order cluster integral, and $b_1=1$.
 We start with exponentiating the relation~(\ref{eq:8}). 
On further expanding the exponential in a power series of $z$, and equating the series 
 power by power to the series  given by~(\ref{eq:02}),
\begin{equation}
{\cal Z}(\beta, z)=1+z Z_1(\beta)+z^2 Z_2(\beta)+z^3 Z_3(\beta)+\dots,
\label{eq:9}
\end{equation}
we obtain, after some algebra,
\begin{equation}
Z_N(\beta)=\frac{1}{N} \sum_{k=1}^{N} kb_k Z_1(\beta) Z_{N-k}(\beta)~,
\label{inter1}
\end{equation} 
where 
\begin{equation}
b_k=b_k^{0}+\Delta b_k~.
\label{step}
\end{equation}
 In the above, $b_k^{0}$ is the statistical, and $\Delta b_k$ the
interaction, part of the $k^{th}$ order virial coefficient.
For ideal (noninteracting) bosons, replacing $b_k$ by ${b_k^{0}}$ in 
(\ref{inter1}), we obtain (\ref{class}), with 
\begin{equation}
b_k^{0}=\frac{1}{k} \frac{Z_1(k\beta)}{Z_1(\beta)}~.
\label{all}
\end{equation}  
With these temperature-dependent virial coefficients in (\ref{inter1}) we recover the exact canonical partition function for the ideal bose system.
We show the $\beta$ dependence of $b_k^0(\beta)$ for ideal boson systems in Fig.~\ref{fig:02}.  The curves have zero slope at $\beta=0$, since in the high-temperature limit $b_k^0=1/k^4$.

\subsection{Fisher zeros for ideal bosons}

We first find  the complex zeros of the 
canonical partition function $Z_N(\beta)$ for noninteracting bosons   
in an isotropic three-dimensional harmonic oscillator. 
 For a fixed $N$, the number of zeros on the 
complex plane, denoted by $\beta_r$, is finite. Therefore $Z_N(\beta) $ may be 
expressed as a finite product, as in (\ref{fish})
\cite{fisher65}. 
For ideal bosons, we use the recursion relation~(\ref{class}).  
 The calculation of $Z_N(\beta)$ for the exact case may be simplified
 by introducing~\cite{schmidt99} $y = e^{\ts -\beta\hbar\omega}$ so that
\begin{equation}\label{eq:a04}
Z_N(y) = \dfrac{y^{3N/2}}{\prod_{j=1}^N(1-y^j)^3}P_N(y),
\end{equation}
with  $P_0(y)=P_1(y)=1$.
$P_N(y)$ is a polynomial in $y$ and when it is zero so is $Z_N(y)$.
Since $P_N(y)$ is a polynomial the number of zeros is equal to its
degree which increases rapidly with particle number.  For example, for
$N=50$ ideal bosons, there are 3495 zeros. 
We determine a subset to 
give a clear indication that the pattern of zeros pinches the positive real $\beta$ axis.
For the continuum
approximation, by contrast, one gets a polynomial in $\beta$, and 
for $N$ ideal bosons, the number of zeros is  $3\times (N-1)$.  

In Fig.~\ref{fig:01} we display the Fisher zeros on the complex $\beta$ plane
for $N=100$ and $N=50$ ideal bosons, using the exact $Z_1(\beta)$
given by (\ref{kish}). 
Even for $N=50$, there is a tendency for the zeros to approach
the real-$\beta$ axis, signalling condensation. 
For $N=100$ ideal bosons, this tendency is more pronounced.   
For an ideal bose gas in an isotropic 3-dimensional harmonic
oscillator, the BEC condensation temperature is given by~\cite{brack03}
\begin{equation}
k_B T_c^{0} \simeq (0.94 N^{1/3}-0.69)~.
\label{id}
\end{equation}
This yields, for $N=50$, $\beta_c=0.36$, and for $N=100$, 
$\beta_c=0.27$, where we have put $k_B=1$. 
These are in agreement with the estimates from Fig.~\ref{fig:01}. 

\begin{figure}[!htb]
\centering
\caption{The Fisher zeros for systems of 50 or 100 noninteracting trapped bosons with a continuous density of states.
}
\label{fig:03}\resizebox{3.5in}{!}{\includegraphics[angle=-90]{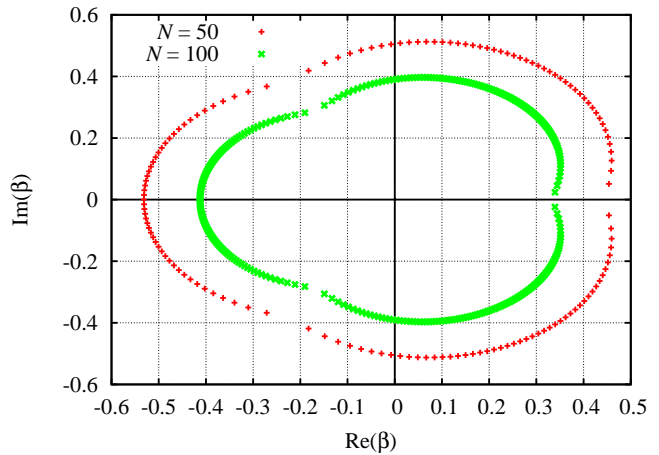}} 
\end{figure}

Note that the statistical virial coefficients, given by
(\ref{all}), are now temperature dependent. Only at high
temperatures do they become temperature independent, as we shall show.
We have verified numerically that we get identical Fisher zeros of the
ideal Bose gas by using (\ref{inter1}), with $b_k$ replaced by
$b_k^{(0)}$, but retaining the proper temperature dependence of the
statistical coefficients. This is of course obvious analytically, but
an important first step in introducing interactions. 
In the high temperature limit, calculations are much easier if one
uses $Z_1(\beta)=1/\beta^3$, which is the
leading term of the exact  $Z_1(\beta)$ given by (\ref{kish}). This corresponds to a 
continuous single-particle density of states that grows quadratically with energy.
Note from (\ref{all}) that in this approximation, the statistical
virial coefficients are temperature independent, given by $b_k^{0}=1/k^{4}$. 
 The Fisher zeros are very sensitive to the analytical properties of
the partition function, so it is interesting to see how different  
the pattern of zeros with this approximation is. 
In Fig.~\ref{fig:03}, we show the pattern of complex zeros of the corresponding
$Z_N(\beta)$ for $N=50$ and $N=100$. Comparison with Fig.~\ref{fig:01} shows 
considerable difference in detail:  the number of zeros being much smaller for
the case of continuous density of states. Nevertheless, both give
agreement  on the tendency of BEC setting in, although  $\beta_c$ is
about $10$ percent higher for the continuous spectrum. 
In both, the estimated condensation temperature  $T_c$ increases
approximately as $N^{1/3}$.

\subsection{Effect of unitary interaction}
To include the effect of interaction, we use (\ref{inter1}). We
consider an attractive zero-range interaction of unitary strength that
gives a zero-energy bound state in the s-wave. We first take the exact $Z_1(\beta)$
given by (\ref{kish}). To calculate $Z_N(\beta)$, it is necessary
to know the {\it exact temperature dependence} of $b_k=b_k^{0}+\Delta b_k$. 
While this is known for the statistical part $b_k^{0}$ exactly, for
the interaction part only the second virial coefficient is known
analytically~\cite{castin13_fr}. The third virial coefficient is complicated
by the Efimov channel, and  given analytically only at the high and
low temperature ends~\cite{castin13_fr}. We therefore  include only the second order
cluster for the interaction part, which for bosons is given by 
\begin{equation}
\Delta b_2=\frac{1}{2}\frac{1}{\cosh (\hbar\omega\beta/2)}~.
\label{inter}
\end{equation}
For the statistical coefficients $b_k^{0}$, all up to $k=N$ are
included.  Since the third virial coefficient is not included, we cannot comment on the effect of the Efimov trimer formations on the Fisher zeros.

\begin{figure}[!hb]
\centering                                                                                      
\caption{The Fisher zeros of systems of 30 bosons with discrete energy spectrum.  Not all the zeros are shown.}
\label{fig:04}
$\begin{array}{c@{\hspace{0.0in}}c@{\hspace{0.0in}}}                                                                                                             
\subfigure[~Zeros of an ideal system of 30 bosons.  There are 1250 zeros.]{
\resizebox{3.5in}{!}{\includegraphics[angle=-90]{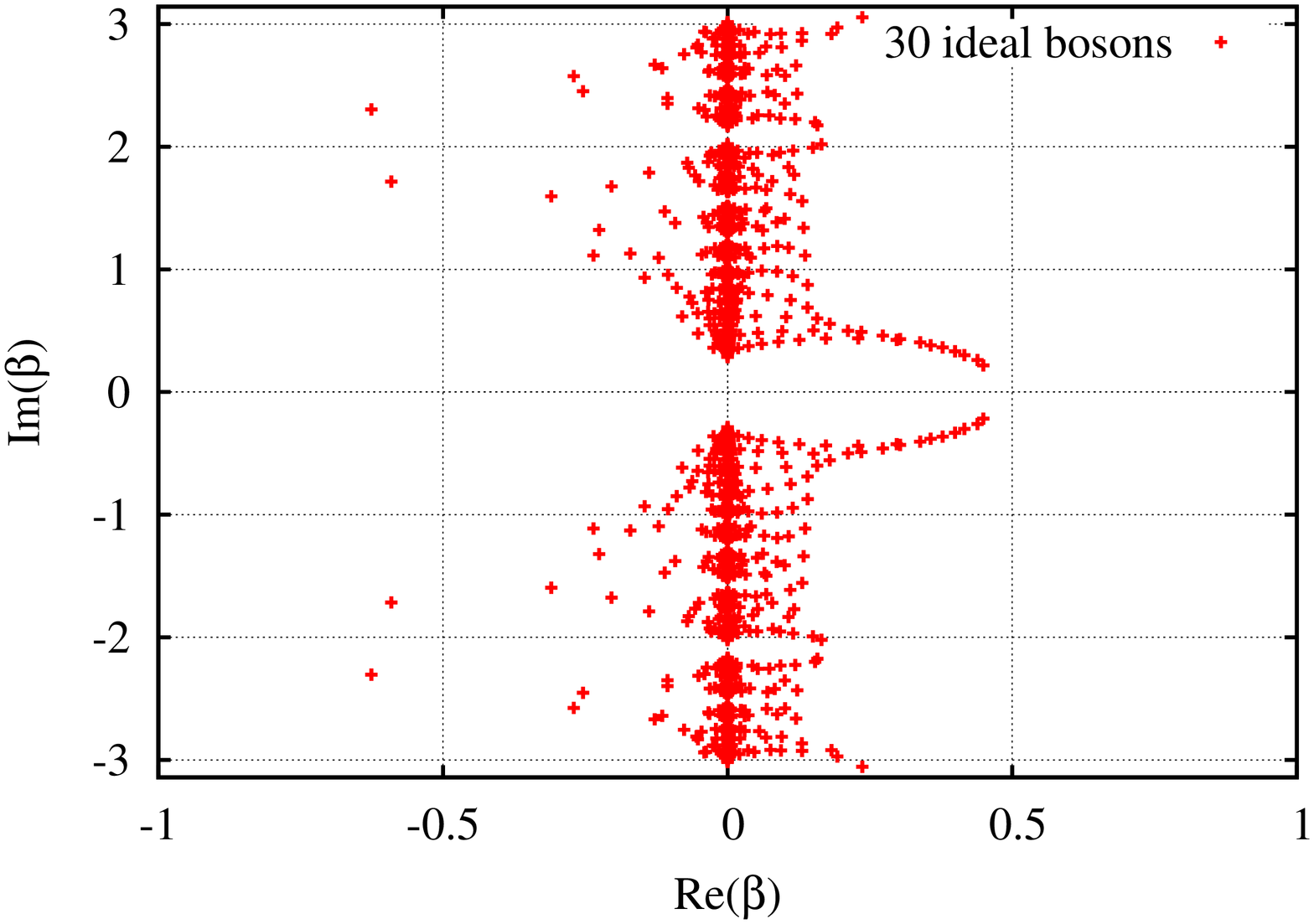}}
\label{fig:04a}} \\ 
\subfigure[~Zeros of unitary system of 30 bosons.  There are 1335 zeros. ]{
\resizebox{3.5in}{!}{\includegraphics[angle=-90]{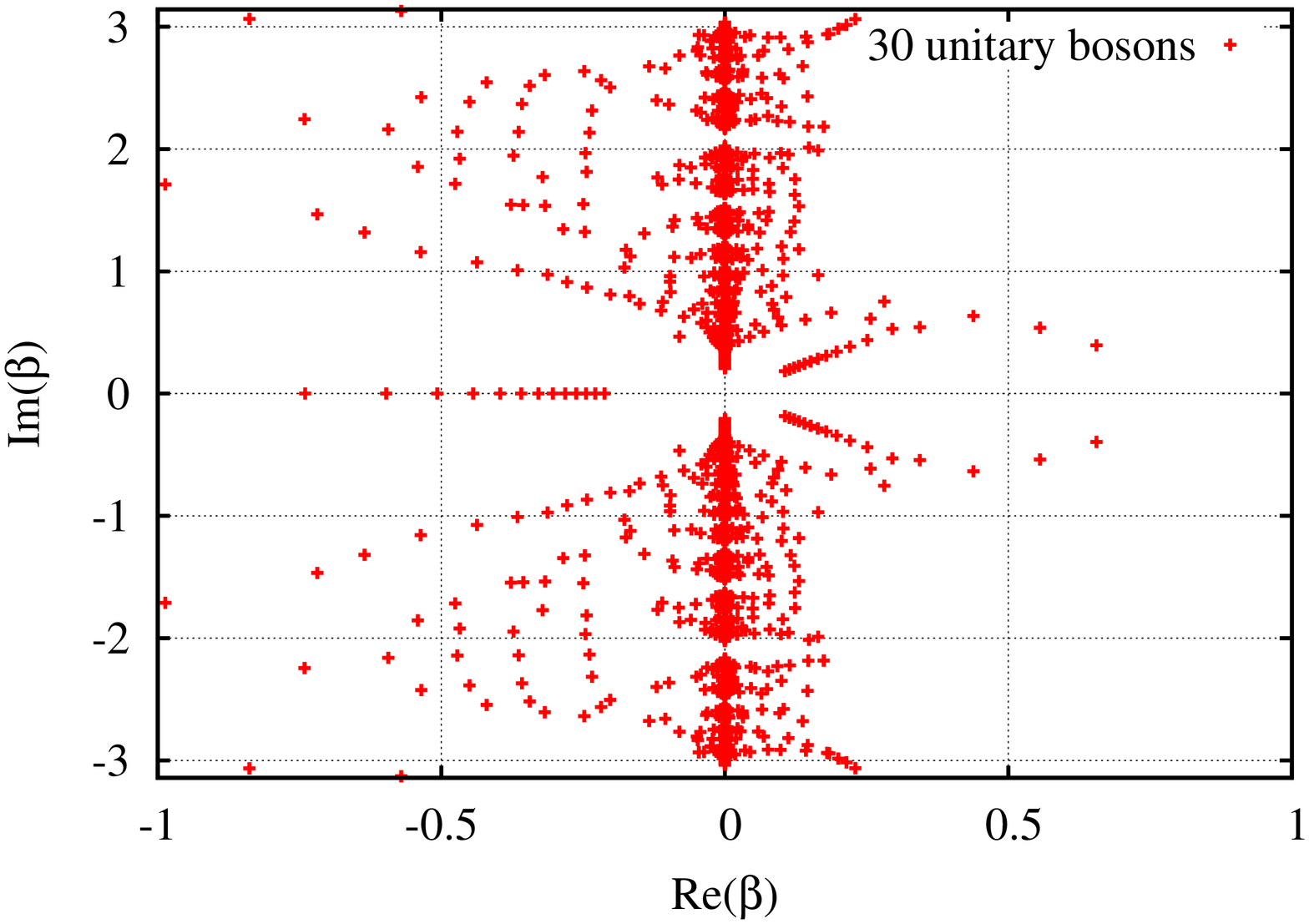}}
\label{fig:04b}}
\end{array}$
\end{figure}

\begin{figure}[h]
\centering                                                                                      
\caption{The Fisher zeros of systems of 100 bosons with continuous  energy spectrum for an ideal and an unitary system.  In both cases there are 297 zeros.}
\label{fig:05}
\resizebox{3.5in}{!}{\includegraphics[angle=-90]{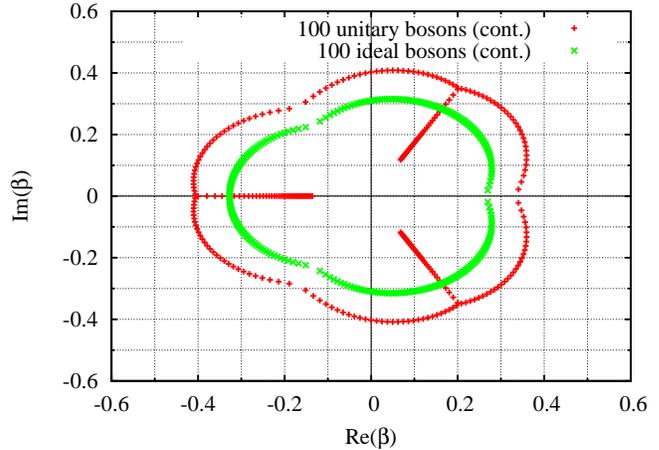}}
\end{figure}

For $30$ bosons, Fig.~\ref{fig:04} displays the Fisher zeros without
and with the unitary interaction. The exact $Z_1(\beta)$ given by
(\ref{kish}) is used. 
Figure~\ref{fig:05} shows a similar comparison for $N=50$ bosons and $N=100$ bosons, respectively, without and with the unitary interaction, but in the continuum approximation for
 $Z_1(\beta)$.   In the exact calculation, the  number of zeros increases
 rapidly with $N$. From Figs.~\ref{fig:04} and \ref{fig:05}, we note that the precursor to BEC
persists even with the attractive unitary interaction. Moreover,
despite some marked differences, 
there is a similarity in the overall patterns of the Fisher
zeros of the ideal and the interacting bosons. In comparison with the 
ideal bosons, the (projected) condensation temperature $T_c$ is considerably lower  
for the interacting one. We surmise that even with the strong unitary
attractive interaction, BEC persists, but one has to go to lower temperatures.
The shift in the critical temperature is considerably larger in our
case than in ~\cite{piatecki14}. This could be because we do not include 
$\Delta b_3$, and/or because $N$ is too small.

\subsection{Specific heat and critical temperature}

It is important to check whether thermodynamic properties
  calculated directly from $Z_N(\beta)$, are consistent with what we
  found from the behaviour of the complex zeros near the critical temperature. We note that it is straightforward to obtain $\langle E\rangle$ and the heat  capacity per particle from the following relations
\begin{equation}\label{eq:15}
\langle E\rangle = - \dfrac{\partial~}{\partial\beta}\ln Z_N(\beta) \mathrm{~~~and~~~} \dfrac{C}{N} = \dfrac{1}{N} \dfrac{\partial\langle E\rangle}{\partial T}.
\end{equation}
\begin{figure}[h!]
\centering                                                                                      
\caption{The specific heat per particle for an ideal and a unitary system of 300 bosons.}
\label{fig:06}
\resizebox{3.5in}{!}{\includegraphics[angle=-90]{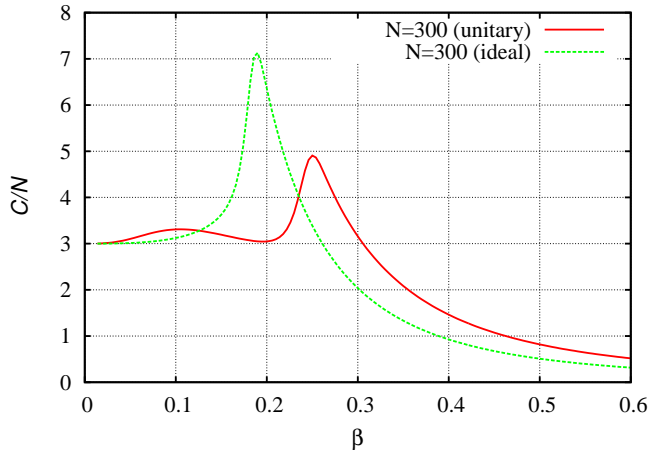}}
\end{figure}
In Fig.~\ref{fig:06} we show the specific heat as a function of temperature for the system with $N=300$ obtained from the exact partition function for both the ideal and unitary cases.  Only the interacting part of the second virial coefficient (\ref{inter})  is included in the unitary system.
Unlike the calculation for the Fisher zeros, the specific heat calculation for a system of a larger number of particle can be done readily.  Figure~\ref{fig:04b} shows for the unitary system a tendency of the pattern of zeros to pinch on the real axis, indicating a critical temperature.  This behaviour becomes more pronounced as the number of particles increases.   In the case of the specific heat the maximum of the curve does not show until a larger number of particles is included, e.g., $N=300$ in Fig.~\ref{fig:06}.  The peak of the curve becomes sharper and higher as $N$ increases further.  The fact that the critical temperature for a unitary system is lower than that of  the ideal system is consistent with the pattern of Fisher zeros.  We also show the specific heat graphs for the continuous-spectrum case of 100-particle system in Fig.~\ref{fig:07}.
\begin{figure}[H]
\centering                                                                                      
\caption{The specific heat per particle for an ideal and a unitary system of 100 bosons with the continuous spectrum (high temperature) approximation.}
\label{fig:07}
\resizebox{3.5in}{!}{\includegraphics[angle=-90]{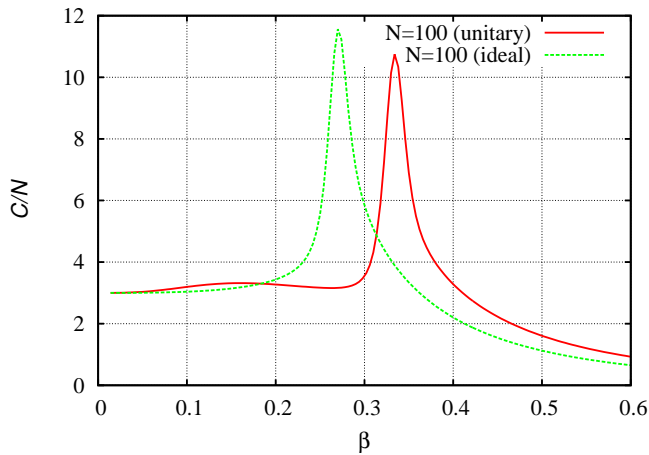}}
\end{figure}
In this case the critical temperatures as read from Fig.~\ref{fig:05} are consistent with the temperatures at which the peaks occur in the graph of the specific heat.

\section{Concluding Remarks}
Fisher zeros for ideal and interacting bosons are examined. Since the
zeros appear all over the complex $\beta$ plane, it is important to 
take the correct temperature dependence of $Z_1(\beta)$, and the
virial coefficients. Only the second virial coefficient as a
function of $\beta$ is analytically known. 
Even then, it is numerically difficult to compute the Fisher zeros for a
large number of bosons.     
Nevertheless, the Fisher zeros of ideal bosons 
even for a very small number (like $N=30$) show a tendency 
towards BEC by  approaching the real $\beta$ axis. This tendency persists, especially as the number of particles increases,  
even when a strong attractive interaction of unitary strength
is introduced through the second virial coefficient.  Remarkably, the high
temperature approximation with a continuous density of states gives 
the same qualitative results. The condensation temperature is
projected to a lower value for the strongly attractive interacting
case, consistent with a direct calculation of the heat capacity 
from $Z_N(\beta)$.  
 Our conclusions are subject
to caution, since the number of bosons are taken to be small, and the 
interacting part of the third and higher virial coefficients are not
included.  For those reasons we do not expect the numerical value of $T_c$ to be accurate. 
Even so, we demonstrate that the behaviour of the complex zeros 
of the partition function of the unitary gas is strongly correlated with
its thermodynamic properties on the real axis.

\section*{Acknowledgements}
The authors are grateful to Professor Akira Suzuki for helpful discussions.


\end{document}